# Reduction of noise and bias in randomly sampled power spectra

Preben Buchhave[1] and Clara M. Velte[2]
1. Intarsia Optics, Birkerød, Denmark
2. Technical University of Denmark, Lyngby, Denmark

## Abstract

We consider the origin of noise and distortions in power spectral estimates of randomly sampled data, specifically velocity data measured with a burst-mode laser Doppler anemometer. The analysis guides us to new ways of reducing noise and removing spectral bias, e.g. distortions caused by modifications of the ideal Poisson sample rate caused by dead time effects and correlations between velocity and sample rate. The noise and dead time effects for finite records are shown to tend to previous results for infinite time records and ensemble averages. For finite records we show that the measured sampling function can be used to correct the spectra for noise and dead time effects by a deconvolution process. We also describe a novel version of a power spectral estimator based on a fast slotted autocovariance algorithm.

## Introduction

The estimation of power spectra of a fluctuating process from a record of randomly sampled data has been the subject of numerous investigations in the past R. B. Blackman & J. W. Tukey (1958) and H.S. Shapiro, R.A. Silverman (1960) and reviews by Durst et al. (1976) and Albrecht et al. (2003). The advantages and disadvantages are well known: When the sampling is sufficiently random (e.g. a Poisson process) and uncorrelated with the process being sampled, it is possible to compute unbiased, alias-free power spectral estimates. However, the randomness in the sampling process induces additional variance in the estimate requiring more data or more sampling time than estimates based on comparable regularly sampled data. Various power spectral estimators have been applied to the problem of laser Doppler measurements of turbulence spectra, and much research has been devoted to a discussion of the advantages and disadvantages of different methods without yet reaching a fully agreed consensus. The so-called direct method of estimating the power spectrum (PS) directly from a modified periodogram was pioneered by Gaster and Roberts in the early 70ties (1977). The condition for aliasing-free spectra was further investigated by Masry (1978). The method applied to LDA processors was investigated by Roberts and Gaster (1980) and Roberts et al. (1980). Residence time weighted PS algorithms that removed the velocity-sample rate correlation was introduced by Buchhave et al. (1979). The alternative method of first estimating the autocovariance and subsequently obtaining the PS by a Fourier transform was pioneered by Mayo (1974) and later developed by Gaster and Roberts (1975). The early estimators suffered from large variance at high frequencies, but this problem was later attacked by many studies, see e.g. reviews by Tropea (1995) and by Benedict, Nobach and Tropea (2000). Of the many methods to improve the slotted autocovariance method (SACF) we may mention the use of variable slot width Tummers and Passchier (1996), fuzzy slot width Nobach (1998) and van Maanen et al. (1999), techniques for varying the slot Benedict et al. (2000), local time estimation for the slotted correlation function by Nobach (2002) and a study of the influence of slot width and particle rate by Benak et al. (1993). Another widely used technique is the so-called sample-and-hold method, which interpolates between velocity records. Already Adrian (1987) showed that a zeroth order interpolation, where a measured velocity value is simply resampled until a new measurement arrives, imparts a severe filtering effect on the computed spectrum. Since then a number of different interpolation schemes have been investigated, and reconstructions by correcting with the filter function have been attempted e.g. Simon (2004) and Moreau (2011). Although these methods have contributed greatly to the art of computing power spectral estimates from LDA data, we in this paper will use only the basic estimators for the direct method and the SACF-method. The reason is that we want to use the raw sampling function for correcting the spectra for sample rate related spectral bias. Any filtering or smoothing procedures are likely to modify the sampling function and thus prevent its use for correction schemes. In fact, a computed spectrum can be no better than allowed by the raw information in the signal. Attempts to improve beyond that is likely to tell more about the data processing than about the underlying power spectrum.

The theory is often treated in an idealized way; the samples are assumed point processes (delta function samples), and the results are based on either infinitely long records or on ensemble averages over infinitely many records. We want to take as our starting point a finite record of a finite number of measured points as it occurs in real measurements. Moreover, we want to include realistic properties such as random sampling noise and the effects of finite width sampling pulses and instrumental dead time that may severely bias spectral estimates.

An important purpose of this paper is to investigate the possibility of correcting the power spectrum computed from the measured data by methods of deconvolution. The final power spectrum is often a convolution of the real spectrum with various spectral filters originating from effects that alter the sampling rate or add noise. We analyze various effects in time delay space and attempt to correct the spectrum by dividing the measured autocovariance function with the autocovariance of the sampling process. Deconvolution is manifestly difficult, especially as regards signals with noise, and different methods are not always successful. We find, however, that it is feasible to remove some of the noise and bias encountered due to non-ideal measurement procedures.

In the following we shall make a more detailed analysis of the noise and bias distorting the power spectra measured with real instruments and try to reduce these effects by e.g. deconvolution methods. Our results apply specifically to the laser Doppler anemometer (LDA) used for the measurement of turbulent velocity power spectra. However, we believe that our methods will be applicable to a wide range of problems. For some of the derivations we refer to recent publications on dead time effects in power spectral estimation (Frontera and Fuligni (1978), Zhang et al. (1995), Buchhave et al. (2014) and Velte et al. (2014b)). However, in the present paper we emphasize the case of realistic measurements with the burst-mode LDA using a finite sample size, and we introduce a term, the noise function, that uniquely describes the randomness associated with a particular record of samples.

## Properties of LDA data

In this publication we shall emphasize the properties of LDA data that affect the estimation of power spectra. We specifically consider the so-called burst-mode LDA, which provides a single velocity data point when a seed particle passes the measurement volume formed by the intersection of two coherent laser beams. We also assume that the processor provides a measurement of the time of arrival and of the time it takes the particle to traverse the measurement volume, which we denote the residence time. We assume that the particles are evenly distributed in the fluid, which would result in a Poisson sampling process if the velocity were constant and the measurement volume were small enough that the probability of more than one particle in the measurement volume at any one time would be negligible. However, when the velocity fluctuates, the sample rate will be correlated with the velocity as more particles are swept through the measurement volume when the velocity is high than when the velocity is low (often referred to as velocity bias). In the following we shall assume that the sample rate is proportional to the instantaneous velocity magnitude at the measurement volume, and we shall assume that the statistical results such as mean velocity, velocity autocovariance function (ACF) and velocity power spectrum (PS) are computed with the so-called residence time weighted (RTW) algorithms that in principle remove velocity bias in the statistical results (George et al. (1978), Buchhave et al. (1979) and Velte et al. (2014a)). An interpretation of this method is to consider the product of the residence time and the velocity as representing a new, unbiased velocity data point that may be used in statistical computations. This quantity will represent the measured velocity data point in the following theoretical analysis, and this will allow us to consider the sampling process and the process being sampled as statistically independent.

The concepts of averaging effect and dead time effect can be demonstrated with reference to Figure 1 below:

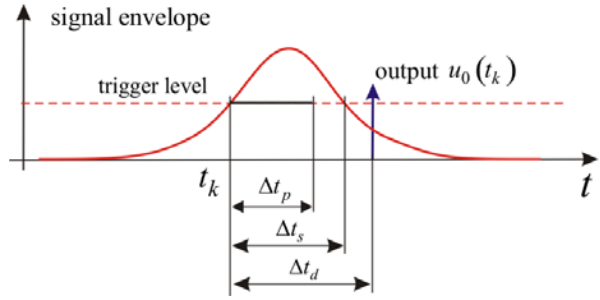

*Figure 1. The sampling process (Buchhave (2014)).*

The figure displays the envelope of a Doppler burst, the electronic signal resulting from the detection of the Doppler modulated light reaching the detector when a particle traverses the measurement volume. The detector signal is amplified and filtered and sent to the signal processor, whose task it is to deduce the Doppler frequency from the modulated burst. When the signal exceeds a certain threshold, the burst detection level, the signal processor will start a fast digitization of the signal, and a subsequent FFT analysis provides, with suitable calibration, a velocity data point. Finally, the resulting velocity data point is transferred to the data processor (or an intermediate buffer memory) for further analysis. The derivation of the Doppler frequency requires a certain number of digital samples and consequently takes a certain time, the processing time $\Delta t_p$, indicated in the figure.

We shall assume that the effect of this processing time is a simple averaging of the real velocity during $\Delta t_p$, which, as we have demonstrated in Buchhave (2014) results in a filtering effect and a reduction of the high frequency part of the spectrum. The duration of the burst, the residence time $\Delta t_s$, corresponds to the maximum possible processing time, and the actual processing time should be adjusted to take into account the minimum $\Delta t_s$ expected in the measurement so that $\Delta t_p$ is always smaller than $\Delta t_s$. The dead time problem occurs because the signal processor is unable to process the signal from a following particle until the burst detector has been reset by the signal going below the burst detector level. The time the signal processor is quenched is called the dead time, $\Delta t_d$, as indicated in the figure. This means that the time between samples and hence the lag used for the computation of the ACF estimate cannot be smaller than the dead time, and the power spectral estimate is distorted by the dead time effect, Buchhave (2014). The situation for LDA measurements of turbulence is complicated partly due to the greatly varying velocity, which influences the measured residence time and partly due to interference that may occur between light scattered from two or more particles being present in the measurement volume at the same time. This complicates the model and computation of the dead time effect, see Velte (2014b) for a detailed analysis. The overall effect of the dead time is an unwanted reduction of the power level at the low end and an oscillation at the high end of the spectrum.

In the following we shall analyze the properties of the random sampling noise and see how the noise, the filtering and the dead time act together to shape the spectrum.

## Sampling function and statistical quantities

The following discussion is general. In subsequent sections we shall investigate the consequences of implementing these results in the numerical recipes for the so-called direct method (DIR) and for the so-called slotted autocovariance method (SACF).

We consider a burst-mode LDA whose function it is obtain data from a stationary turbulent velocity,

$u(t) = \bar{u} + u'(t)$, where $\bar{u}$ is the temporal mean and $u'(t)$ is the fluctuating part. In the following we shall refer to the carrier of the information, the Doppler modulated electronic pulse, as the "signal" and to the result of the measurement, $u_0(t)$, as the "velocity". The total output from the signal processor for each burst is the time of arrival, $t_k$, the residence time, $\Delta t_s$, and the measured velocity, $u_0(t_k)$, see Figure 1.

We assume a sampling function, $g(t-t_k)$, which assigns the measurement to a particular arrival time $t_k$ and describes the detector response and the statistical properties of the sampling process. The measured velocity, $u_0(t)$, can then be written as a continuous function of time (see, e.g., George (1978))

$$u_0(t) = u(t) g(t). \tag{1}$$

We shall further assume that the data is collected in a number $M$ of records or blocks of a finite record length, $T_g$, and that data arrivev with the average sampling rate, $\nu$. The records or blocks of data are selected from independent parts of the total measurement record.

## Ideal sampling

Ideally, a measurement of a time dependent point process should occur at a single point in space and time. We may represent random sampling of such a process by a sampling function, which is a string of delta-functions placed at random arrival times, $t_k$:

$$g(t) = \frac{1}{\nu} \sum_{k=1}^{N} \delta(t - t_k) \tag{2}$$

Note that we have normalized by $\nu$, the average number of samples per unit time, to obtain the correct scaling and dimension of the sampling function $g(t)$. $N$ is the actual number of samples collected in the record. Figure 2 is a sketch of a typical $g(t)$.

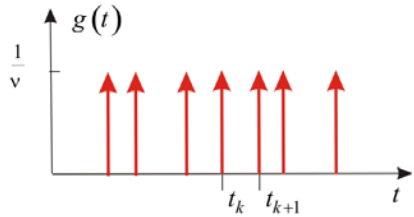

*Figure 2. Ideal random sampling represented by a string of delta functions at random time $t_k$.*

Thus the measured velocity becomes

$$u_0(t) = u(t) \cdot \frac{1}{\nu} \sum_{k=1}^{N} \delta(t - t_k) \qquad 0 \leq t_k < T_g \tag{3}$$

The string of delta functions records the exact arrival times of all the velocity samples in the record.

## Real signals

We consider a randomly sampled signal with both a finite processing time, $\Delta t_p$, which may be equal to the residence time, $\Delta t_s$, but may also be smaller as determined by the properties of the signal processor. We also consider a finite dead time, $\Delta t_d$, which may be as small as $\Delta t_s$ or greater as required by the data transfer. We describe the measurement process by a top hat sampling function of width $\Delta t_p$ and height $1/(\nu \Delta t_p)$, see Figure

3. We assume that the effect of the dead time is that no sample can follow a previous sample within the time lag $\Delta t_d$. This we describe by the condition $t_{k+1} - t_k > \Delta t_d$.

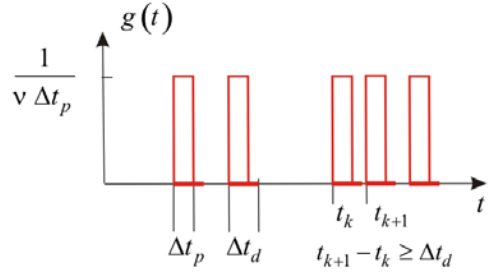

*Figure 3. Top hat sampling function with dead time.*

The top hat sampling function may be represented by

$$g(t) = \frac{1}{\Delta t_p} \sum_{k=1}^{N} \Pi_{\Delta t_p}(t_k) = \frac{1}{\Delta t_p} \sum_{k=1}^{N} \Pi_{\Delta t_p}(t) \frac{1}{v} \delta(t - t_k) = \frac{1}{\Delta t_p} \Pi_{\Delta t_p}(t) \cdot \frac{1}{v} \sum_{k=1}^{N} \delta(t - t_k), \quad (4)$$

where the factor $1/(v \Delta t_p)$ calibrates time averages of the measured signal. $\Pi_{\Delta t_p}(t_k)$ designates a rectangular top hat function of unity height and width $\Delta t_p$, initiated at the arrival time $t = t_k$.

### Signal averaging effects
The detector and signal processor are assumed to cause a simple averaging of the true velocity during the processing time, $\Delta t_p$. We represent this by the convolution of the true velocity and the sampling function. Thus the measured velocity becomes:

$$u_0(t) = \left[ u(t) \otimes \frac{1}{\Delta t_p} \Pi_{\Delta t_p}(t) \right] \cdot \frac{1}{v} \sum_{k=1}^{N} \delta(t - t_k) \quad (5)$$

This can be interpreted as a filtered velocity arriving at random times:

$$u_0(t) = u_{\Delta t_p}(t) \cdot \frac{1}{v} \sum_{k=1}^{N} \delta(t - t_k) \quad (6)$$

where $u_{\Delta t_p}(t) = u(t) \otimes \frac{1}{\Delta t_p} \Pi_{\Delta t_p}(t)$. In the following, the top hat sampled velocity will be treated as a string of delta functions, just like the ideal velocity above, but with a filtered velocity value given by Eq. (6). In Buchhave (2014) we showed that the effect of the top hat sampling function on the measured power spectrum is the product of the true spectrum and a sinc-squared frequency transmission function:

$$S_{0,\Delta t_p}(f) = S_0(f) \cdot \text{sinc}^2(\pi f \Delta t_p) \quad (7)$$

### Dead time effects
In Buchhave (2014) we showed that the effect of a fixed dead time is a bias or distortion of the measured spectrum given by a convolution of the true spectrum with the factor

$$S_{0,\Delta t_p, \Delta t_d}(f) = S_{0,\Delta t_p}(f) \otimes \left[ \delta(f) - 2\Delta t_d \, \text{sinc}(2\pi f \Delta t_d) \right] \quad (8)$$

In a LDA measurement, the dead time will be highly variable due to velocity variations and random particle trajectories (Velte (2014b)). However, as the dead times are presumed measured it will be possible to correct for the effect even in the LDA case as described below. The filtering and dead time affect the velocity before the digitization and calculation of the spectrum. Therefore the dead time effect can be handled in the following discussion of the computed power spectrum by using the modified true velocity $u_{\Delta t_p, \Delta t_d}$, which includes the effects of both top hat filtering and dead time.

We continue the discussion with a velocity sampled by a string of delta functions, but with the understanding that the real sampling function can be handled by inserting the modified velocity $u_{\Delta t_p, \Delta t_d}$.

## Mean value

We assume that the true velocity is a stationary random process and that we measure a record that includes the longest time scales in the flow ($T_g > 2T_i$, where $T_i$ is the integral time scale). Assuming a low level Poisson distribution for the sampling process, the probability of a measurement in $dt$ is

$$\begin{cases} P\{0\}dt \cong 1 \\ P\{1\}dt \cong (1/\bar{N})\nu dt \\ P\{>1\}dt \cong 0 \end{cases} \quad (9)$$

where $\bar{N} = \nu T_g$ is the mean number of samples.

We find for the mean sampling function during the record:

$$\langle g(t) \rangle = \int_{-\infty}^{\infty} P\{1\} \frac{1}{\nu} \sum_{k=1}^{N} \delta(t-t_k) \, dt = \frac{1}{\bar{N}} \sum_{k=1}^{N} \int_{-\infty}^{\infty} \delta(t-t_k) \, dt = \frac{N}{\bar{N}} \cong 1, \quad (10)$$

where we have used the fact that the measured number of samples, $N$, is approximately equal to the mean number of samples, $N \cong \bar{N} = \nu T_g$, for all but very short records.

The mean value of the sampled velocity is then, assuming statistical independence between the measurement process and the sampled process or the use of RTW weighting,

$$\langle u_0(t) \rangle = \langle g(t) \rangle \cdot \langle u(t) \rangle = \bar{u} \quad (11)$$

which is of course equal to the true mean velocity as we have normalized the sampling function.

## Autocovariance function

An estimate for the measured ACF is given by

$$\hat{C}_0(\tau) = \langle u'(t)g(t) \cdot u'(t')g(t') \rangle = \langle u'(t)u'(t') \rangle \langle g(t)g(t') \rangle = C_{u'}(\tau) \cdot \hat{C}_g(\tau) \quad (12)$$

where $\tau = t' - t$ is the time lag and $\hat{C}_g(\tau)$ is the ACF estimate of the sampling function based on a single record of length $T_g$. $C_{u'}(\tau)$ is the ACF of the true fluctuating part of the velocity, and we have again assumed independence between the measurement process and the velocity.

We can visualize the ACF of the sampling function, $\hat{C}_g(\tau)$, by imagining a copy of the delta function record of $N$ samples sliding past the original delta function record with the lag time $\tau$. Correlations will then occur $N^2$ times when the lag matches a measured time between samples, see Figure 4.

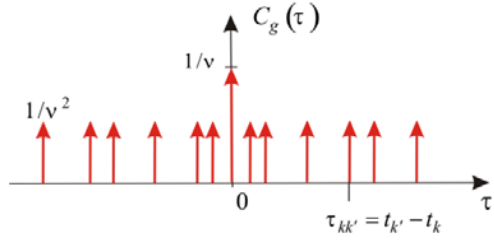

*Figure 4. ACF for a delta function sampling function.*

We can then estimate the sampling function ACF based on a single record by

$$\hat{C}_g(\tau) = \int_{-\infty}^{\infty} P\{1\} \frac{1}{v}\sum_{k=1}^{N}\delta(t-t_k) \frac{1}{v}\sum_{k'=1}^{N}\delta(t-t_{k'}+\tau)\,dt$$

$$= \frac{1}{v\bar{N}} \sum_{k,k'}^{N} \delta(\tau - \tau_{kk'}) \qquad (13)$$

where $\tau_{kk'} \equiv t_k - t_{k'}$ are the possible lags or times-between-samples, and the summation is over all pairs of measured values.

We can write this as a sum of self-product terms ($k=k'$) and cross-product terms ($k \neq k'$):

$$\hat{C}_g(\tau) = \frac{1}{v}\delta(\tau) + \frac{1}{v\bar{N}}\sum_{k\neq k'}^{N}\delta(\tau-\tau_{kk'}). \qquad \left(0 \leq |\tau_{kk'}| < T_g\right) \qquad (14)$$

The second term is a unique "fingerprint" for each measurement and describes the actual lag times available for the measured record. We have coined the term "noise function in time delay space" for this term. The first term is a spike at zero time tag. It is an artifact of the way we compute the ACF and it does not contain any information on the flow.

The ACF estimate for the measured velocity is then

$$\hat{C}_0(\tau) = C_{u'}(\tau)\cdot \hat{C}_g(\tau) = C_{u'}(\tau)\cdot\left[\frac{1}{v}\delta(\tau) + \frac{1}{v\bar{N}}\sum_{k\neq k'}^{N}\delta(\tau-\tau_{kk'})\right] \qquad (15)$$

The ensemble value of the measured ACF is

$$C_0(\tau) = C_{u'}(\tau)\cdot\left[\frac{1}{v}\delta(\tau) + \left\langle\frac{1}{v\bar{N}}\sum_{k\neq k'}^{N}\delta(\tau-\tau_{kk'})\right\rangle\right] \qquad (16)$$

We have assumed a record length $T_g$ so the effective overlap area will taper off linearly, which means that the probability of measuring a covariance sample at $\tau$ will decrease linearly with increasing delay as $W(\tau)d\tau = (1/T_g)(1-|\tau|/T_g)d\tau$ providing a double sided triangular window.

The mean value of cross products is then

$$P\{1\} = \frac{1}{2T_g}\int_{-T_g}^{T_g}\frac{1}{v\bar{N}}\sum_{k\neq k'}^{N}\delta(\tau-\tau_{kk'})d\tau = 1$$

Thus, the ensemble average of the sampling function ACF is simply

$$C_0(\tau) = \frac{1}{v}\delta(\tau) + \left(1-\frac{|\tau|}{T_g}\right) \qquad \left(-T_g < \tau < T_g\right) \qquad (17)$$

For $T_g \rightarrow \infty$ this becomes

$$C_g(\tau) = \frac{1}{\nu}\delta(\tau) + 1 \qquad (-\infty < \tau < \infty) \qquad (18)$$

and

$$C_0(\tau) = C_{u'}(\tau) \cdot \left(\frac{1}{\nu}\delta(\tau) + 1\right) \qquad (-\infty < \tau < \infty) \qquad (19)$$

as presented in e.g. Velte (2014a) and references herin.

## Power spectrum using the Wiener-Khintchine theorem

The power spectrum estimate can be found by using the Wiener-Khintchine theorem (the autocovariance method):

$$\hat{S}_0(f) = \mathrm{FT}\{\hat{C}_0(\tau)\} = \mathrm{FT}\{C_{u'}(\tau) \cdot \hat{C}_g(\tau)\} = \mathrm{FT}\{C_{u'}(\tau)\} \otimes \mathrm{FT}\{\hat{C}_g(\tau)\}$$
$$= S_{u'}(f) \otimes \hat{S}_g(f) \qquad (20)$$

where $S_{u'}(f)$ is the true spectrum, and $\hat{S}_g(f)$ is the Fourier transform of the sampling function ACF:

$$\hat{S}_g(f) = \int_{-\infty}^{\infty} e^{-i2\pi f\tau} \left[\frac{1}{\nu \bar{N}} \sum_{k,k'}^{N} \delta(\tau - \tau_{kk'})\right] d\tau$$
$$= \frac{1}{\nu \bar{N}} \sum_{k,k'}^{N} e^{-i2\pi f\tau_{kk'}} \qquad (21)$$

Thus

$$\hat{S}_0(f) = S_{u'}(f) \otimes \frac{1}{\nu \bar{N}} \sum_{k,k'}^{N} e^{-i2\pi f\tau_{kk'}} \qquad (22)$$

We can again break this up into self-products and cross-products:

$$\hat{S}_0(f) = \frac{\overline{u'^2}}{\nu} + S_{u'}(f) \otimes \frac{1}{\nu \bar{N}} \sum_{k \neq k'}^{N} e^{-i2\pi f\tau_{kk'}} \qquad (23)$$

where we have again used that $N \cong \nu T_g = \bar{N}$. The second term we have denoted the "noise function in frequency space". The equation shows how the self-product term is reduced relative to the cross-product term as the average sample rate increases or equivalently, as the number of samples in the measured record increases.

We can now include the effects of top hat sampling and dead time:

$$\hat{S}_{0,\Delta t_p,\Delta t_d}(f) = \frac{\overline{u^2_{\Delta t_p,\Delta t_d}}}{\nu_0} + S_{u',\Delta t_p,\Delta t_d}(f) \otimes \frac{1}{\nu \bar{N}} \sum_{k \neq k'}^{N} e^{-i2\pi f\tau_{kk'}} \qquad (24)$$

or

$$\hat{S}_{0,\Delta t_p,\Delta t_d}(f) = \frac{\overline{u^2_{\Delta t_p,\Delta t_d}}}{\nu_0} + \left[S_{u',\Delta t_p}(f) \otimes \left[\delta(f) - 2\Delta t_d \,\mathrm{sinc}(2\pi f\Delta t_d)\right]\right] \otimes \frac{1}{\nu \bar{N}} \sum_{k \neq k'}^{N} e^{-i2\pi f\tau_{kk'}} \qquad (25)$$

or explicitly

$$\hat{S}_{0,\Delta t_p,\Delta t_d}(f) = \frac{\overline{u^2_{\Delta t_p,\Delta t_d}}}{\nu_0}$$
$$+ \left[\left[S_{u'}(f) \cdot \mathrm{sinc}^2(\pi f\Delta t_p)\right] \otimes \left[\delta(f) - 2\Delta t_d \,\mathrm{sinc}(2\pi f\Delta t_d)\right]\right] \otimes \frac{1}{\nu \bar{N}} \sum_{k \neq k'}^{N} e^{-i2\pi f\tau_{kk'}} \qquad (26)$$

where $\nu_0$ is the sample rate reduced by dead time (Buchhave (2014)). We continue with the simpler expression Eq.(23) realizing that we can always include the effects of filtering and dead time as in Eq.(26). Using

$$\left\langle \frac{1}{v\bar{N}} \sum_{k \neq k'}^{N} e^{-i2\pi f \tau_{kk'}} \right\rangle = \mathrm{sinc}^2 \left( 2\pi f T_g \right), \text{ where } |\tau_{kk'}| \text{ is uniformly distributed in the range } -T_g \leq \tau_{kk'} \leq T_g, \text{ we get}$$

the measured spectrum for a finite record length $T_g$

$$S_0(f) = \frac{\overline{u'^2}}{v} + S_{u'}(f) \otimes \mathrm{sinc}^2 \left( 2\pi f T_g \right). \tag{27}$$

and for an infinite record

$$S_0(f) = \frac{\overline{u'^2}}{v} + S_{u'}(f) \tag{28}$$

Note: The sinc-squared window requires a uniform distribution of $\tau_{kk'}$ in the range $-T_g \leq \tau_{kk'} \leq T_g$. A finite record, where $\tau_{kk'}$ tapers off with a triangular window leads to the stronger oscillating $\mathrm{sinc}\left(2\pi f T_g\right)$ spectral window.

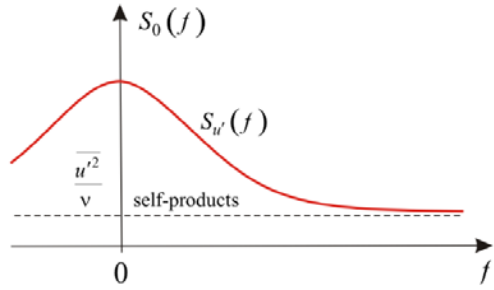

*Figure 5. The measured spectrum consists of the true spectrum and a spectral offset.*

The measured spectrum is identical to the true spectrum plus a constant offset term that does not include any spectral information, see Figure 5. This term is a consequence of the way we calculate the power spectrum and is easily computed and could be subtracted as done by Gaster and Roberts and others, or one could leave it, since in practice there will be other (some possibly frequency dependent) offset terms that have to be dealt with. Another reason for leaving the self-products or part of them in the PS is that log-log plots are often used, and that a suitable offset is just a vertical shift making it easier to evaluate the spectrum.

## Power spectrum by the direct method

The so-called direct method is mathematically equivalent to the above method that used the Wiener-Khinchin theorem. However, the direct method is often computationally more efficient in array based math software and therefore often the preferred method.

The formula for the direct method is:

$$\begin{aligned}
\hat{S}_0(f) &= \lim_{T \to \infty} \frac{1}{T} \tilde{u}'_0(f)^* \tilde{u}'_0(f) \\
&= \lim_{T \to \infty} \frac{1}{T} \int_{-T/2}^{T/2} \int_{-T/2}^{T/2} e^{-i2\pi f (t'-t)} \langle u'(t) u'(t') \rangle \, dt \, dt' \otimes \frac{1}{v\bar{N}} \sum_{k,k'}^{N} e^{-i2\pi f \tau_{kk'}} \\
&= S_{u'}(f) \otimes \frac{1}{v\bar{N}} \sum_{k,k'}^{N} e^{-i2\pi f \tau_{kk'}}
\end{aligned} \tag{29}$$

where $\tilde{u}'_0(f)$ is the Fourier transform of the measured fluctuating velocity and $S_{u'}(f)$ represents the true power spectrum. $\hat{S}_0(f)$ is still a power spectral estimate based on a single record. The result is the true spectrum convolved with the sampling function power spectral estimate for the record. This is the same result as above.

## Spectral variance

The variance is defined as:

$$\text{var}\{S_0(f)\} = \langle \hat{S}_0(f)^2 \rangle - \langle \hat{S}_0(f) \rangle^2 \equiv I - II \tag{30}$$

The second term above is the square of the mean, given by

$$II = \langle \hat{S}_0(f) \rangle^2 = \left[ \frac{\overline{u'^2}}{v} + S_{u'}(f) \otimes \left\langle \frac{1}{v\overline{N}} \sum_{k \neq k'}^{N} e^{-i2\pi f \tau_{kk'}} \right\rangle \right]^2 \tag{31}$$

or

$$II = \left(\frac{\overline{u'^2}}{v}\right)^2 + 2\frac{\overline{u'^2}}{v} S_{u'}(f) \otimes \text{sinc}^2(2\pi f T_g) + S_{u'}(f)^2 \otimes \text{sinc}^4(2\pi f T_g) \tag{32}$$

To find an expression for the first term, we assume that the velocity fluctuations are fourth order Gaussian. We then find for the variance of the measured spectrum, see Appendix A:

$$\text{var}\{S_0(f)\} = \left[ S_{u'}(f) \otimes \left\langle \frac{1}{v} + \frac{1}{v\overline{N}} \sum_{k \neq k'}^{N} e^{(\tau - \tau_{kk'})} \right\rangle \right]^2 \tag{33}$$

or ensemble averaged for finite record length

$$\text{var}\{S_0(f)\} = \left(\frac{\overline{u'^2}}{v}\right)^2 + 2\frac{\overline{u'^2}}{v} S_{u'}(f) \otimes \text{sinc}^2(2\pi f T_g) + S_{u'}(f)^2 \otimes \text{sinc}^4(2\pi f T_g) \tag{34}$$

The convolution with the sinc-squared is just the usual convolution with the frequency window corresponding to a rectangular time window. We could have used another time window like the Hanning window, but the effect on turbulence spectra is minor as long as the record is much longer then the integral time scale and no distinct spectral features are present.

For very long records, $T_g \to \infty$, we find

$$\text{var}\{S_0(f)\} = \left(\frac{\overline{u'^2}}{v}\right)^2 + 2\frac{\overline{u'^2}}{v} S_{u'}(f) \otimes \delta(f) + S_{u'}(f)^2 \otimes \delta(f)^2$$

$$= \left[ \left(\frac{\overline{u'^2}}{v}\right) + S_{u'}(f) \right]^2 \tag{35}$$

This result is in agreement with e.g. Gaster and Roberts (1975).

## Random sampling noise

### Frequency content of the noise

To see the effect of the randomness in the sampling process, consider again Eq.(23):

$$\hat{S}_0(f) = S_{u'}(f) \otimes \frac{1}{v\overline{N}} \sum_{k,k'}^{N} e^{-i2\pi f \tau_{kk'}} = \frac{\overline{u'^2}}{v} + S_{u'}(f) \otimes \frac{1}{v\overline{N}} \sum_{k \neq k'}^{N} e^{-i2\pi f \tau_{kk'}} \tag{36}$$

The second term is a convolution between the true spectrum and a zero-mean noise term consisting of a sum of exponentials with random phases. The values of $\tau_{kk'}$ cause a unique and deterministic fluctuation or sampling noise on the spectral estimate related to the particular measurement. If the record length is finite, the effect of the convolution is a broadening of the real spectrum in analogy with the finite resolution obtained with a finite record of regularly sampled data.

Finally, we note that the noise term includes the convolution with a possible mean velocity

$$S_u(f) \otimes \frac{1}{\nu \overline{N}} \sum_{k \neq k'}^{N} e^{-i2\pi f \tau_{kk'}} = \left[\overline{u}^2 \delta(f) + S_{u'}(f)\right] \otimes \frac{1}{\nu N} \sum_{k \neq k'}^{N} e^{-i2\pi f \tau_{kk'}} \tag{37}$$

It is thus advisable to subtract the mean before the power spectral estimator is applied. The spectral estimates in this paper are all computed for the fluctuating part of the velocity.

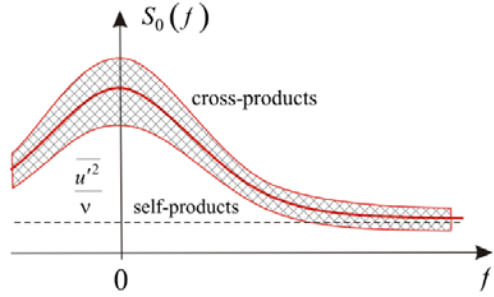

*Figure 6. Total measured spectrum (solid line), frequency dependent noise (cross hatched region) reflecting the spectral distribution due to finite number of samples and the constant offset (broken line).*

As can be seen by inspection of the convolution integral of the cross terms or the noise function,

$$S_{u'}(f) \otimes \frac{1}{\nu \overline{N}} \sum_{k \neq k'}^{N} e^{-i2\pi f \tau_{kk'}} = \int_{-\infty}^{\infty} S_{u'}(f') \frac{1}{\nu \overline{N}} \sum_{k \neq k'}^{N} e^{-i2\pi (f-f')\tau_{kk'}} df', \tag{38}$$

the frequency spectrum of the noise reflects the spectral distribution of the velocity (the noise is greater where the true spectrum is high). The cross hatched area in Figure 6 illustrates the noise in the power spectral estimate due to the finite number of samples. The convolution of the noise function with the spectrum means that a wide spectrum gives a relatively low frequency noise and a sharp spectral line would lead to high frequency noise. As the sample rate increases, but the record length remains finite, the noise is reduced and the spectrum is ultimately convolved with the sinc-squared frequency function caused by a rectangular time window.

Figure 7 shows the power spectrum of a spectral line computed by the direct method based on a single record with a high sample rate (left) and with a lower sample rate (center). In case of the true spectrum having a finite spectral content (right), we see the effect of the convolution of the true spectrum and the noise function on the spectral content of the noise. In all three cases the self-products are included resulting in a spectral offset. In agreement with the expression for the variance, the high frequency part of the spectrum (outside the influence of the true spectrum) consists of two parts, the constant spectral offset and a zero mean fluctuation caused by the randomness of the sample arrival times. Notice that the spectrum including the self-products never goes below zero whereas a spectrum with the self-products subtracted, due to the random sampling noise, would fluctuate around zero at frequencies where the true spectrum is zero.

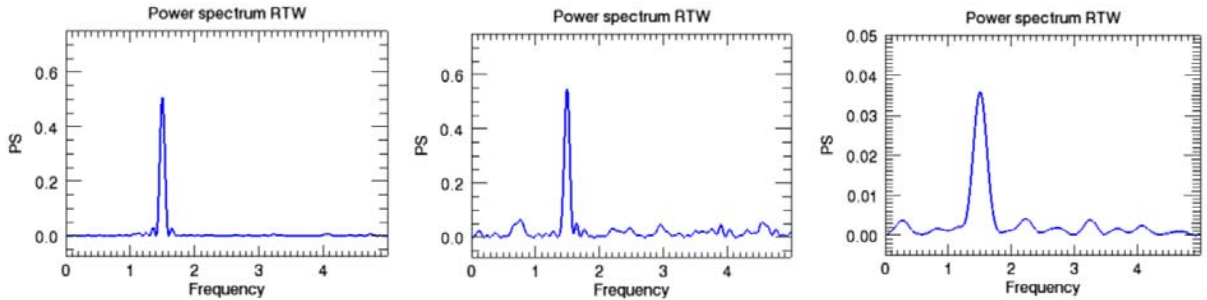

*Figure 7. Power spectrum of a finite record of computer generated velocity data. Left: Narrow spectral line, high sample rate, Center: Narrow spectral line, low sample rate. Right: Broad spectral line, low sample rate.*

# Correcting the power spectrum by deconvolution

## Theory

### Correcting for filtering

As we have seen above, several processes contribute to the distortion of the measured power spectrum. We have argued that the fact that a velocity measurement requires a certain time, the processing time, $\Delta t_p$, for the digitizing and processing of the Doppler burst results in the multiplication of the true spectrum by a sinc-squared transfer function. Knowing $\Delta t_p$ it should be possible to restore the estimate of the true spectrum by a simple division in frequency space by the transfer function:

$$\hat{S}_0(f) = \hat{S}_{0,\Delta t_p}(f) / \operatorname{sinc}^2(\pi f \Delta t_p) \tag{39}$$

However, this effect is generally much smaller that the effect of dead time.

### Correcting for dead time

The effect of a possible dead time is a convolution in frequency space of the true spectrum (corrected for the averaging effect) with a dead time function as shown in Eq.(26). Restoration would require division in correlation space by a known dead time ACF. This process would be rather straightforward in case of a fixed dead time (Buchhave (2014)), but is rendered more complex in case of real LDA data by the variations in residence time (Velte (2014b)).

However, as we shall see, measurement of the sampling function ACF allows us to compensate the PS for any process that modifies the sample rate, including noise and the dead time effect.

### Correcting for random sampling noise

As shown in Eq.(24), the random sampling noise can be described as a convolution of the estimate of the true spectrum modified by filtering and dead time with a noise function characteristic of the particular record. In correlation space, the noise enters as a product of the velocity ACF and a noise ACF. A reconstruction, which would remove the noise, would involve the division of the measured velocity ACF by the noise function ACF:

$$\hat{C}_{0,correced}(\tau) = \hat{C}_0(\tau) / \hat{C}_g(\tau). \tag{40}$$

Fortunately, the arrival times are known from the measured record and can be used for deconvolution.

### Deconvolution by means of the measured sampling function ACF

The detrimental effects mentioned above can be corrected by means of the measured sampling function ACF. Referring to Eq. (26), the filtered spectrum, $S_{0,\Delta t_p}(f)$, is convolved with the dead time function and the noise function

$$S_{0,\Delta t_p}(f) \otimes \left[\delta(f) - 2\Delta t_d \operatorname{sinc}(2\pi f \Delta t_d)\right] \otimes \frac{1}{\nu \bar{N}} \sum_{k \neq k'}^{N} e^{-i2\pi f \tau_{kk'}} \tag{41}$$

In correlation space this is the product of three ACFs:

$$\hat{C}_{0,\Delta t_p}(\tau) \cdot \left[1 - \Pi_{\Delta t_d}(\tau)\right] \cdot \hat{C}_g(\tau) \equiv \hat{C}_{0,\Delta t_p}(\tau) \cdot \hat{C}_g^M(\tau) \tag{42}$$

We may call $\hat{C}_g^M(\tau)$ the measured sampling function ACF. Thus, we can obtain the spectrum corrected for both noise and dead time by division in correlation space by the measured sampling function ACF:

$$\hat{C}_u(\tau) = \hat{C}_0(\tau) / \hat{C}_g^M(\tau). \tag{43}$$

In fact, the measured sampling function ACF will correct for any effect that modifies the sampling rate, for example dead time effects, electronic filtering effects and random sampling. Velocity bias is fully taken into account by the residence time weighting.

However, the measured sampling function ACF, $\hat{C}_g^M(\tau)$, is a double sum of delta functions (Eq.(13)), and it is not obvious how this division could be performed. In the following we shall describe two methods of implementing Eq.(43). As expected, the effects of noise and dead time show up differently for different spectral estimators.

## Implementation of deconvolution

There are numerous ways to estimate the power spectrum when it comes to the practical programming. We shall describe deconvolution applied to two of these methods in the following, the direct method and the slotted autocovariance method. We shall verify the results by means of computer generated velocity data and by application of the methods to measured velocity data obtained in a turbulent free jet in air.

The computer generated data has been described in Buchhave (2014). Briefly, the random samples were grabbed from a high rate primary velocity record with a Von Karman power spectrum by a Poisson process. The Von Karman spectrum was chosen to model as closely as possible the jet spectrum described below.

$$S_{vK}(f) = \frac{1}{62.5} \cdot \frac{1}{\left(1 + (f/45)^2\right)^{5/6}} \cdot \exp\left(-(f/2500)^{4/3}\right). \tag{44}$$

The integral time scale of this process can be found from $l = 2\,\text{var}(u)/S(0)$.

The Poisson process was modulated by the velocity magnitude to include velocity bias, but otherwise adjusted to give only zeros or ones (and a very low number of twos). Additional (phase) noise was not included in this presentation although it is included in the computer programs. If included, it only adds to the noise floor of the spectrum.

The jet measurements were described in Velte (2014a).

### *The direct method with computer generated data*

In the first method, we compute the power spectrum through a DFT with equidistant frequencies on both the random velocity samples and the random sampling function. We then do an inverse FFT on these spectra to compute the corresponding ACFs. Then we perform the division indicated in Eq.(43) and finally convert back to frequency space by an FFT process. The process is repeated for $M$ records, and the final spectrum is found by block averaging.

The problem with this method is that because of the random sampling noise, the sampling function ACF may include zero values or negative values making the division unstable. The common method to circumvent this problem is to use a Wiener deconvolution, where the mean square noise is added in the denominator, or simply to add a small, constant value to $C_g(\tau)$. (As it turns out, the constant is not really necessary when many records are block averaged. The infinities and negative spectral values are smoothed out by the averaging process). By this process, we obtain the power spectrum of a set of computer generated data shown in Figure 8. The input parameters for the data for this plot were: Mean velocity 5 ms$^{-1}$, turbulence intensity 25%, record length: 1 s, number of records for block averaging: 200, average sample rate 4500 s$^{-1}$, measurement volume diameter: 40 μm.

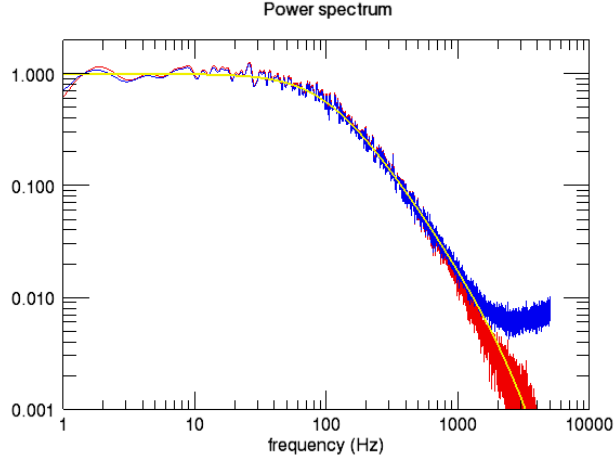

*Figure 8. Power spectrum of computer generated velocity data. Yellow: Von Karman model spectrum, blue: direct method, red: same power spectrum after applying deconvolution. Both averaged over 200 records.*

The figure shows three curves: The yellow curve shows the model Von Karman spectrum forming the basis for the computer generated data. The blue curve shows the power spectrum computed by the direct method and the red curve shows the result of the deconvolution described above. The compensation for the dead time effect is evident. The noise may seem to increase by the deconvolution, but here we must remember that we are looking at a logarithmic plot, where fluctuations look larger at low values of the spectrum. The noise is further reduced by block averaging (200 records). The red and the blue spectra are based on exactly the same original data.

### *The SACF method with computer generated data*

In the second method, we compute directly the SACF and find the power spectrum by FFT of the measured ACF. However, the most straightforward method to do this is also prohibitively slow. In that method we compute all possible lags between samples in a record. This provides $N^2$ possible values for the delay $\tau_{kk'}$, if the record consists of $N$ samples. We then have to sort the products of the corresponding velocities into $n_{slot}$ slots covering the total range of time delays, where $n_{slot}$ must be high enough to avoid aliasing and also allow us to measure the lowest frequency we want, and finally we must divide the sum of products in each slot by the number of measurements in the slot. In order to compute a spectrum up to high frequencies and to avoid introduction of aliasing, the slot width $\Delta\tau$ must be smaller than $1/(2f_N)$, where $f_N$ is the desired Nyquist frequency for the spectrum Velte (2014a). The small $\Delta\tau$ means a large number of slots in order to also cover the longest time scales in the flow, and sorting of the double sum into the slots and performing all the velocity products will require much computing time.

Actually, in the RTW method we normalize with the product of residence times Velte (2014a):

$$\hat{C}_0(m\Delta\tau) = \frac{\sum_{k,k'}^{N} u_k \Delta t_k u_{k'} \Delta t_{k'}}{\sum_{k,k'}^{N} \Delta t_k \Delta t_{k'}} \quad \text{with} \quad (m-\tfrac{1}{2})\Delta\tau < \tau_{kk'} < (m+\tfrac{1}{2})\Delta\tau \quad \text{and} \quad \tau_{kk'} \equiv t_{k'} - t_k \quad (45)$$

Add to this that we average $M$ spectra, and we get a huge computational load. Instead we developed an alternative, where we sort the measured velocities into $n_{slot}$ arrival time slots and perform the ACF by multiplying the velocity record with its delayed copy. This is the method one would use for regularly sampled data, but in our case there will be a large number of empty slots due to the randomness in the arrival times and the small amount of data compared to the number of slots. In the software it is possible to skip multiplications when a slot is empty, and we can compute the ACF for each record. We then loop the program $M$ times and form the average of $M$ ACFs. This method turns out to be much faster than the conventional method described above. We have compared the dynamic range and noise level for the two methods, the "double sum method", where all $N^2$ possible lags between samples in a record have been used, and the new method, which only uses a one dimensional array of $N$ lags. We have found that the computed spectra are very similar, in fact the single sum spectra tend to have a little less noise. The reason for this is again the fact that the $N$ samples in a record contain all the available information and nothing is gained by reusing it with the $N^2$ possible lags.

Figure 9 compares the direct method and the novel SACF method for a situation with negligible dead time. The computational parameters were: Mean velocity 5 ms$^{-1}$, turbulence intensity 25%, record length: 1 s, number of frequencies estimated, direct method 1000, SACF 50.000, number of slots in SACF: 100.000, average data rate approx. 4500 s$^{-1}$, number of records for block averaging: 400, measurement volume 0.01 µm. The yellow curve is again the von Karman spectrum behind the computer-generated data. Evidently, there is a perfect agreement between the two different methods of computing the power spectrum. The measurement volume is assumed so small that no dead time effect will be present in this case, and the processing time $\Delta t_p$ is assumed insignificant.

Note: No additional corrections beyond the basic estimators were made.

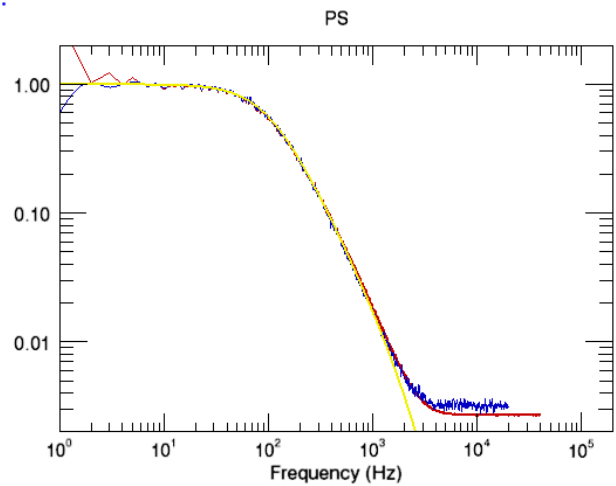

*Figure 9. Power spectrum of computer generated velocity data with small measurement volume. Yellow: Von Karman model spectrum, blue: direct method, red: slotted autocovariance method. Spectral offset subtracted. Both averaged over 400 records.*

We now argue that the expression Eq. (43) performs the desired deconvolution and compensates for both noise and dead time. The sum in the accumulator is simply the sum of residence time weighted velocity products and the denominator is the total overlap time for the measured velocity pairs with arrival time difference falling within the slot with time delay $m\Delta\tau$. Thus $\hat{C}_g^M(m\Delta\tau)$ includes all effects that modify the sampling rate, and the division indicated in Eq.(43) compensates for all of these effects. Thus it can be said that the power spectrum computed by the residence time weighted slotted ACF method compensates for both velocity bias, dead time and the random sampling noise by deconvolution. Likewise, the averaging or normalization of velocity products performed by the conventional SACF-methods described in the references, which is known to reduce noise in the SACF-estimate, can be viewed as a deconvolution, process reducing both noise and sample rate modifying errors.

Figure 10 shows again a comparison between the direct method and the SACF method, but in this case with a larger measurement volume (simulation parameters: mean velocity 5 ms$^{-1}$, turbulence intensity 25%, record length 1 s, measurement volume 40 μm diameter, averages sample rate approx. 3500 s$^{-1}$, number of records: 200). The blue curve shows the power spectrum evaluated by the direct method. Dead time effects show up by the oscillation (the "dip" at high frequencies). Also, the spectral offset is higher than in the previous figure as a result of the lower data rate (the offset is not subtracted). The red curve shows the SACF power spectrum estimated from the same data set with the same data rate and number of records. The dead time effect on the spectrum is removed by the deconvolution implicit in the method. (The slight dip can perhaps be attributed to a dead time effect due to the minimum lag given by the slot width. Also, the SACF cannot compensate if the slot width is greater than the dead time so that the variation in the dead time cannot be resolved).

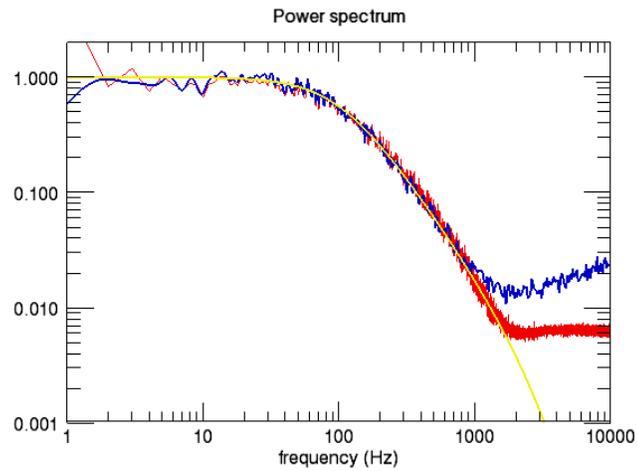

*Figure 10. Power spectrum of computer generated velocity data with finite measurement volume. Yellow: Von Karman model spectrum, blue: direct method, red: slotted autocovariance method. Block averaged 200 records.*

## Measurements

Figure 11 illustrates deconvolution applied to data measured in a turbulent jet. The experimental parameters for this plot is: Jet exit diameter: 10 mm, measurement location: 30 diameters downstream, 26 mm off axis, mean velocity: 3.1 ms$^{-1}$, turbulence intensity: 48 %, Re = 20000, integral time scale 0.0048 s. The active measurement volume was 75X700 μm and the average data rate was approx. 5000 s$^{-1}$. The scales are estimated as: Kolmogorov scale = 53 μm, Taylor scale = 2.2 mm at the center line. 4.000.000 data points were used.

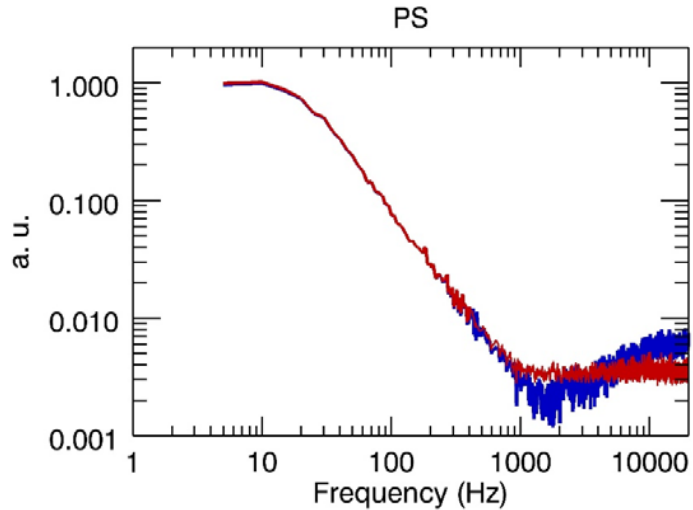

*Figure 11. Power spectrum of measured velocity data in a turbulent jet in air. Blue: direct method, red: direct method with deconvolution. 800 record block average.*

Figure 12 shows the same data processed by the SACF method. The blue curve is again the direct method. The red curve is the SACF-method applied to the same data. The yellow curve shows the effect of averaging the red curve by a range of neighboring frequency estimates. The range of frequencies averaged is increased from 5 at the low frequency end to 100 at the high frequency end. The logarithmic plot causes a great number of frequency estimates to be concentrated at the high frequency end, and since the spectrum does not change much in this range it is permissible (and indeed advantageous) to average a greater range of spectral estimates at the high frequency end.

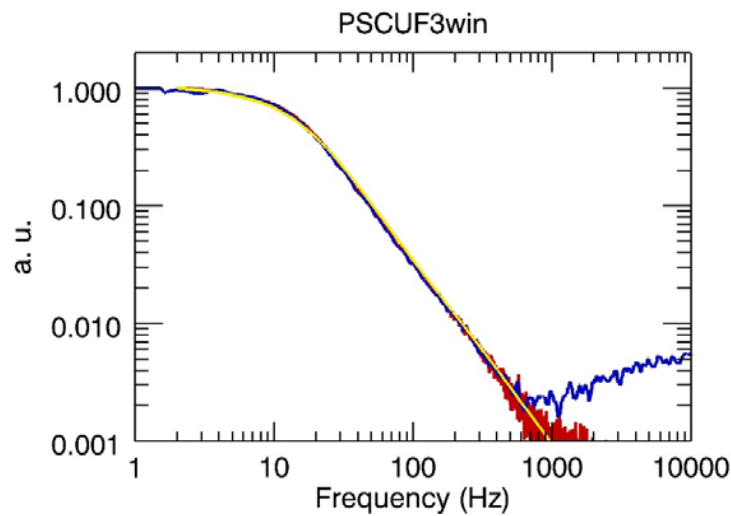

*Figure 12. Power spectrum of measured velocity data in a turbulent jet in air. Blue: direct method, red: SACF method, yellow: SACF with nonlinear averaging. 800 record block average.*

## Conclusions

We have described randomly sampled data from a burst-mode LDA by means of a realistic sampling function that allows description of various effects that occur in a real measurement. These include averaging due to the finite processing time needed to produce a velocity measurement, dead time effects due to the inability of the processor to produce data with arbitrarily short time between samples and a function we denoted the noise function, which contains the noise information in either time delay space or frequency space for a particular record. We further showed that the measured power spectral estimate of a single randomly measured velocity record can be presented as a convolution of the true power spectrum, a "dead time function" that accounts for the dead time effect and a spectral noise function. Corresponding to the convolution in frequency space, a product of the true velocity ACF, a dead time ACF and an ACF noise function exists in time delay space. The product of the dead time ACF and the noise ACF is precisely the ACF of the measured sampling function constructed from the measured arrival times. Thus a deconvolution of the measured power spectrum is possible by division of the measured velocity ACF with the measured sampling function ACF in time delay space. The corrected measured power spectrum is then obtained by a Fourier transform of the corrected measured ACF. Other instrument related effects such as filtering or quantization of the numerical values may affect the sampling rate, and we postulate that the convolution of the dead time function and the noise function can be considered as parts of a more general sampling function that accounts for all effects that distort the sample rate and thereby induce bias in the measured power spectrum.

We tested this idea with two different spectral estimators, the direct method and the slotted autocovariance method. We implemented the slotted autocovariance with a novel algorithm, which greatly reduces the computation time and still obtains spectra of the same quality as previous algorithms.

The effect of the deconvolution performed on realistic computer generated data is promising. The effect of dead time is essentially removed and the useful dynamic range until noise drowns out the spectrum is improved significantly in both examples.

# Appendix A

## Variance

The variance of the measured power spectrum is defined as

$$\text{var}\{S_0(f)\} = \langle \hat{S}_0(f)^2 \rangle - \langle \hat{S}_0(f) \rangle^2 \equiv I - II, \qquad \text{A1}$$

where subscripts $0$ stands for "measured with a record length $T_g$", respectively, and the circumflex (hat) indicates estimate from a single record. In the following we distinguish between a time average parameter $T$ used to compute the true spectrum of the continuous velocity and the actual record length $T_g$ determined by the measured samples, see Fig. A1.

Let us look first at the second term above, $II$.

### *Second term (square of the mean):*

We understand $\hat{S}_0(f)$ as an estimate of the power spectrum based on a finite record of length $T_g$, see Figure A1. We then get, using the direct method and the expression for the noise function power spectrum derived above:

$$\hat{S}_0(f) = \lim_{T \to \infty} \frac{1}{T} \tilde{u}_0(f)^* \tilde{u}_0(f) \qquad \text{A2}$$

$$= \lim_{T \to \infty} \frac{1}{T} \int_{-T/2}^{T/2} dt \int_{-T/2}^{T/2} dt' e^{-i2\pi f(t'-t)} u'(t) u'(t') \otimes \frac{1}{\nu \bar{N}} \sum_{k,k'}^{N} e^{-i2\pi f \tau_{kk'}} \qquad \text{A3}$$

In the limit of $T \to \infty$, the first term is the true spectrum. The convolution with the second term imposes the finite record on the result and generates the discrete version of the estimator:

$$\hat{S}_0(f) = \frac{1}{\nu \bar{N}} \sum_{k,k'}^{N} e^{-i2\pi f(t_{k'}-t_k)} u'(t_k) u'(t_{k'}) \qquad \text{A4}$$

The ensemble mean of the spectral estimate is then

$$S_0(f) = S_{u'}(f) \otimes \frac{1}{\nu \bar{N}} \left\langle \sum_{k,k'}^{N} e^{-i2\pi f \tau_{kk'}} \right\rangle \equiv S_{u'}(f) \otimes S_g(f) \qquad \text{A5}$$

where $\tau_{kk'} \equiv t_k - t_{k'}$ are the measured delays between the record and a delayed copy.

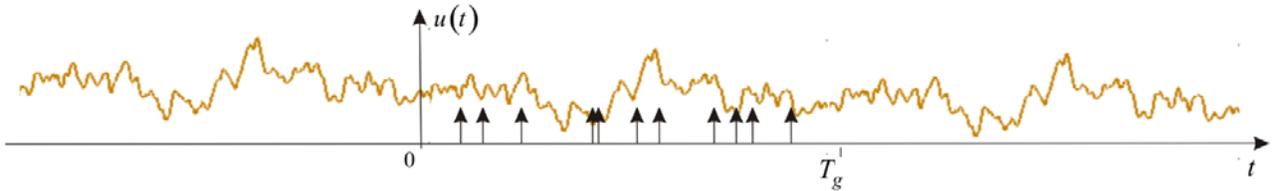

*Figure A1. The true velocity $u(t)$ and a measurement with record length $T_g$ made by sampling function $g(t)$.*

Then the second term above is

$$II = \left[ S_{u'}(f) \otimes S_g(f) \right]^2 \qquad \text{A6}$$

*First term (mean square):*

We understand $\langle \hat{S}_0(f)^2 \rangle$ as the ensemble average of the product of two copies of the same measured power spectrum. By taking the Fourier transform of four copies of the same velocity record sampled at $t, t', t'', t'''$ spanning the same time interval $\{-T/2, T/2\}$ where we let $T$ go to infinity, we obtain:

$$\langle \hat{S}_0(f)^2 \rangle = \lim_{T\to\infty} \frac{1}{T^2} \left\langle \int_{-T/2}^{T/2}\int_{-T/2}^{T/2}\int_{-T/2}^{T/2}\int_{-T/2}^{T/2} \left[ dt\,dt'\,dt''\,dt'''\, e^{-i2\pi f[(t'-t)+(t'''-t'')]} \right.\right.$$
$$\left.\left. \cdot \left[ u'(t)\frac{1}{\nu}\sum_{k}^{N}\delta(t-t_k)u'(t')\frac{1}{\nu}\sum_{k'}^{N}\delta(t'-t_{k'})u'(t'')\frac{1}{\nu}\sum_{k''}^{N}\delta(t''-t_{k''})u'(t''')\frac{1}{\nu}\sum_{k'''}^{N}\delta(t'''-t_{k'''}) \right] \right]\right\rangle$$

A7

or, rearranging and using independence between velocity and sampling,

$$\langle \hat{S}_0(f)^2 \rangle = \lim_{T\to\infty} \frac{1}{T^2} \int_{-T/2}^{T/2}\int_{-T/2}^{T/2}\int_{-T/2}^{T/2}\int_{-T/2}^{T/2} \left[ dt\,dt'\,dt''\,dt'''\, e^{-i2\pi f[(t'-t)+(t'''-t'')]} \langle u'(t)u'(t')u'(t'')u'(t''') \rangle \right.$$
$$\left. \cdot \frac{1}{\nu^4}\left\langle \sum_{k,k',k'',k'''}^{N} \delta(t-t_k)\delta(t'-t_{k'})\delta(t''-t_{k''})\delta(t'''-t_{k'''}) \right\rangle \right]$$

A8

The can be expressed as a convolution of the velocity part and the sampling function part:

$$\langle \hat{S}_0(f)^2 \rangle = \lim_{T\to\infty} \frac{1}{T^2} \int_{-T/2}^{T/2}\int_{-T/2}^{T/2}\int_{-T/2}^{T/2}\int_{-T/2}^{T/2} dt\,dt'\,dt''\,dt'''\, e^{-i2\pi f[(t'-t)+(t'''-t'')]} \langle u'(t)u'(t')u'(t'')u'(t''') \rangle$$

$$\otimes \lim_{T\to\infty} \frac{1}{\nu^2 \bar{N}^2} \int_{-T/2}^{T/2}\int_{-T/2}^{T/2}\int_{-T/2}^{T/2}\int_{-T/2}^{T/2} dt\,dt'\,dt''\,dt'''\, e^{-i2\pi f[(t'-t)+(t'''-t'')]} \left\langle \sum_{k,k',k'',k'''}^{N} \delta(t-t_k)\delta(t'-t_{k'})\delta(t''-t_{k''})\delta(t'''-t_{k'''}) \right\rangle$$

A9

**Velocity part**

To proceed with the velocity part we assume fourth order Gaussian statistics:

$$\lim_{T\to\infty} \frac{1}{T^2} \int_{-T/2}^{T/2}\int_{-T/2}^{T/2}\int_{-T/2}^{T/2}\int_{-T/2}^{T/2} dt\,dt'\,dt''\,dt'''\, e^{-i2\pi f[(t'-t)+(t'''-t'')]}$$
$$\cdot \left[ \langle u'(t)u'(t') \rangle \cdot \langle u'(t'')u'(t''') \rangle + \langle u'(t)u'(t'') \rangle \cdot \langle u'(t')u'(t''') \rangle + \langle u'(t)u'(t''') \rangle \cdot \langle u'(t')u'(t'') \rangle \right]$$

A10

*First velocity term in mean square*

$$\boxed{1} = \lim_{T\to\infty} \frac{1}{T^2} \int_{-T/2}^{T/2}\int_{-T/2}^{T/2}\int_{-T/2}^{T/2}\int_{-T/2}^{T/2} dt\,dt'\,dt''\,dt'''\, e^{-i2\pi f[(t'-t)+(t'''-t'')]} \langle u'(t)u'(t') \rangle \langle u'(t'')u'(t''') \rangle$$

A11

Split into product of two integrals

$$\boxed{1} = \lim_{T\to\infty} \frac{1}{T} \int_{-T/2}^{T/2}\int_{-T/2}^{T/2} e^{-i2\pi f(t'-t)} \langle u'(t)u'(t')\rangle dt\,dt' \cdot \lim_{T\to\infty} \frac{1}{T} \int_{-T/2}^{T/2}\int_{-T/2}^{T/2} e^{-i2\pi f(t'''-t'')} \langle u'(t'')u'(t''')\rangle dt''\,dt''' \qquad \text{A12}$$

$$\boxed{1} = S_{u'}(f)^2 \qquad \text{A13}$$

### Second velocity term in mean square

$$\boxed{2} = \lim_{T\to\infty} \frac{1}{T^2} \int_{-T/2}^{T/2}\int_{-T/2}^{T/2}\int_{-T/2}^{T/2}\int_{-T/2}^{T/2} dt\,dt'\,dt''\,dt'''\, e^{-i2\pi f\left[(t'-t)+(t'''-t'')\right]} \langle u'(t)u'(t'')\rangle\langle u'(t')u'(t''')\rangle \qquad \text{A14}$$

Shuffle variable in exponent to match covariance terms:

$$\boxed{2} = \lim_{T\to\infty} \frac{1}{T^2} \int_{-T/2}^{T/2}\int_{-T/2}^{T/2}\int_{-T/2}^{T/2}\int_{-T/2}^{T/2} dt\,dt'\,dt''\,dt'''\, e^{-i2\pi f\left[(t''-t)+(t'''-t')+2t'-2t''\right]} \langle u'(t)u'(t'')\rangle\langle u'(t')u'(t''')\rangle \qquad \text{A15}$$

We then again have a product of two independent integrals, but with additional phase shifts:

$$\boxed{2} = \lim_{T\to\infty} \frac{1}{T} \int_{-T/2}^{T/2}\int_{-T/2}^{T/2} e^{+i4\pi f t''} e^{-i2\pi f(t''-t)} \langle u'(t)u'(t'')\rangle dt\,dt'' \cdot \lim_{T\to\infty} \frac{1}{T} \int_{-T/2}^{T/2}\int_{-T/2}^{T/2} e^{+i4\pi f t'} e^{-i2\pi f(t'''-t')} \langle u'(t')u'(t''')\rangle dt'\,dt'''$$

A16

Introducing $\tau'' \equiv t''-t$ and $\tau''' \equiv t'''-t'$ we get

$$\boxed{2} = \lim_{T\to\infty} \int_0^T e^{+i4\pi f t''} dt'' \int_0^T e^{-i2\pi f \tau''} C_{u'}(\tau'') d\tau'' \cdot \lim_{T\to\infty} \int_0^T e^{+i4\pi f t'} dt' \int_0^T e^{-i2\pi f \tau'''} C_{u'}(\tau''') d\tau''' \qquad \text{A17}$$

$$\boxed{2} = \left[\delta(f)\cdot S_{u'}(f)\right]^2 = 0 \qquad \text{A18}$$

### Third velocity term in mean square

$$\boxed{3} = \lim_{T\to\infty} \frac{1}{T^2} \int_{-T/2}^{T/2}\int_{-T/2}^{T/2}\int_{-T/2}^{T/2}\int_{-T/2}^{T/2} dt\,dt'\,dt''\,dt'''\, e^{-i2\pi f\left[(t'-t)+(t'''-t'')\right]} \langle u'(t)u'(t''')\rangle\langle u'(t')u'(t'')\rangle \qquad \text{A19}$$

Shuffle variable in exponent:

$$\boxed{3} = \lim_{T\to\infty} \frac{1}{T^2} \int_{-T/2}^{T/2}\int_{-T/2}^{T/2}\int_{-T/2}^{T/2}\int_{-T/2}^{T/2} dt\,dt'\,dt''\,dt'''\, e^{-i2\pi f\left[(t'''-t)+(t'-t'')\right]} \langle u'(t)u'(t''')\rangle\langle u'(t')u'(t'')\rangle \qquad \text{A20}$$

We then again have a product of two independent integrals:

$$\boxed{3} = S_{u'}(f)^2 \qquad \text{A21}$$

Thus the sum of the three terms becomes

$$\boxed{1} + \boxed{2} + \boxed{3} = 2S_{u'}(f)^2 \qquad \text{A22}$$

Subtracting the square of the mean, we get for the velocity part of the variance:

$$\text{var}_{velocity}\{S_0(f)\} = S_{u'}(f)^2 ,\qquad\text{A23}$$

which is the standard expression for the variance of the power spectrum of a stationary random variable.

**Sampling function part**

We can now consider the expression for the sampling function:

$$\frac{1}{\nu^2 \overline{N}^2} \lim_{T\to\infty} \int_{-T/2}^{T/2}\int_{-T/2}^{T/2}\int_{-T/2}^{T/2}\int_{-T/2}^{T/2} dt\, dt'\, dt''\, dt'''\, e^{-i2\pi f\left[(t'-t)+(t'''-t'')\right]} \sum_{k,k',k'',k'''}^{N} \delta(t-t_k)\delta(t'-t_{k'})\delta(t''-t_{k''})\delta(t'''-t_{k'''}) \qquad\text{A24}$$

Using the delta functions one by one, we simply get:

$$\frac{1}{\nu^2 \overline{N}^2} \sum_{k,k',k'',k'''}^{N} e^{-i2\pi f\left[(t_{k'}-t_k)+(t_{k'''}-t_{k''})\right]} \qquad\text{A25}$$

or, defining $\tau_{kk'} \equiv t_{k'} - t_k$ and $\tau_{k''k'''} \equiv t_{k'''} - t_{k''}$,

$$\frac{1}{\nu^2 \overline{N}^2} \sum_{k,k',k'',k'''}^{N} e^{-i2\pi f\left[\tau_{kk'}+\tau_{k''k'''}\right]} \qquad\text{A26}$$

We can now see how the sampling function picks out individual velocity samples through the convolution with the exponentials.

Since the samples are uncorrelated (Poisson sampling), we can write

$$\frac{1}{\nu^2 \overline{N}^2} \sum_{k,k',k'',k'''}^{N} e^{-i2\pi f\left[\tau_{kk'}+\tau_{k''k'''}\right]} = \left(\frac{1}{\nu \overline{N}} \sum_{k,k'}^{N} e^{-i2\pi f \tau_{kk'}}\right)^2 \qquad\text{A27}$$

**Final result**

Thus, for the total variance

$$\text{var}\{\hat{S}_0(f)\} = \lim_{T\to\infty} \int_{-T/2}^{T/2}\int_{-T/2}^{T/2}\int_{-T/2}^{T/2}\int_{-T/2}^{T/2} dt\, dt'\, dt''\, dt''' \, e^{-i2\pi f\left[(t'-t)+(t'''-t'')\right]} \langle u'(t)u'(t')u'(t'')u'(t''')\rangle$$

$$\otimes \left\langle \frac{1}{\nu \overline{N}} \sum_{k,k'}^{N} e^{-i2\pi f \tau_{kk'}} \right\rangle^2 \qquad\text{A28}$$

$$= \left[ S_{u'}(f) \otimes \left\langle \frac{1}{\nu \overline{N}} \sum_{k,k'}^{N} e^{-i2\pi f \tau_{kk'}} \right\rangle \right]^2 \qquad\text{A29}$$

$$= \left[ \frac{\overline{u'^2}}{v} + S_{u'}(f) \otimes \left\langle \frac{1}{v\bar{N}} \sum_{k \neq k'}^{N} e^{-i2\pi f \tau_{kk'}} \right\rangle \right]^2 \quad \text{A30}$$

$$= \left[ \frac{\overline{u'^2}}{v} + S_{u'}(f) \otimes \text{sinc}^2(2\pi f T_g) \right]^2 \quad \text{A31}$$

$$= \left( \frac{\overline{u'^2}}{v} \right)^2 + 2\frac{\overline{u'^2}}{v} S_{u'}(f) \otimes \text{sinc}^2(2\pi f T_g) + S_{u'}(f)^2 \otimes \text{sinc}^4(2\pi f T_g) \quad \text{A32}$$

The convolution with the sinc-squared is just the usual convolution with the frequency window corresponding to a rectangular time window $\{0, T_g\}$.

For an infinite record, $T_g \to \infty$, we find

$$\text{var}\{S_0(f)\} = \left( \frac{\overline{u'^2}}{v} \right)^2 + 2\frac{\overline{u'^2}}{v} S_{u'}(f) + S_{u'}(f)^2 = \left[ \frac{\overline{u'^2}}{v} + S_{u'}(f) \right]^2. \quad \text{A33}$$